\documentclass[10pt,letterpaper]{article}
\usepackage{opex3}
\usepackage{cite}
\begin{document}
\title{
Role of the Coulomb potential on the ellipticity in atomic high-order harmonics generation
}
\author{Xiaosong Zhu,$^{1}$ Meiyan Qin,$^{1}$ Qingbin Zhang,$^{1,2,3}$ Weiyi Hong,$^{1,2}$ Zhizhan Xu,$^{1}$ and Peixiang Lu$^{1,2}$$^{\star}$}
\address{$^1$ Wuhan National Laboratory for Optoelectronics and School of Physics, Huazhong University of
Science and Technology, Wuhan 430074, China\\
$^2$ Key Laboratory of Fundamental Physical Quantities Measurement of Ministry of Education, Wuhan 430074, China\\
$^3$ zhangqingbin@mail.hust.edu.cn, China}
\email{$^\star$ lupeixiang@mail.hust.edu.cn}

\begin{abstract}
The role of the Coulomb potential on the generation of elliptically polarized high-order harmonics from atoms driven by elliptically polarized laser is investigated analytically. It is found that the Coulomb effect makes a contribution to the harmonic ellipticity for low harmonic orders and short quantum path. By using the strong-field eikonal-Volkov approximation, we analyze the influence of the Coulomb potential on the dynamics of the continuum state and find that the obtained harmonic ellipticity in our simulation originates from the break of symmetry of the continuum state.
\end{abstract}

\section{Introduction}
When atoms and molecules are exposed to intense laser irradiation, high-order harmonics (HH) will be generated. The high-order harmonic generation (HHG) has been an attractive topic for the past two decades for its potential application of producing coherent attosecond pulses \cite{Mairesse,Goulielmakis,Lan,Hong1,Hong2,Strelkov1,Etches1} in the XUV regime and self-probing of molecules allowing a combination of attosecond and {\AA}ngstr\"{o}m resolutions \cite{Lein,Haessle,Itatani,Hijano,Chen}. The physical mechanism of HHG can be understood by the three-step model: (i) an atom or molecule emits its electron to continuous state, (ii) the electron is accelerated in the laser field, (iii) the accelerated electron recombines with the parent ion and high energy photon is radiated \cite{Corkum,Schafer}. As described by this model, linearly polarized high-order harmonics will be generated from atomic gases by linearly polarized laser pulses. It is also found that, when the laser pulses are elliptically polarized, elliptically polarized HH will be obtained \cite{Antoine}. In the observed phenomena, the rotation of the polarization of the HH field with respect to that of the driving laser can be explained by the classical three-step model \cite{Weihe}, but this classical model can not explain the harmonic ellipticity. Quantum models based on the strong-field approximation (SFA) can be employed to provide theoretical approaches to investigate the observed ellipticity \cite{Antoine,Antoine2,Strelkov2}. However, these theoretical approaches did not suggest qualitative explanation for the origin of the ellipticity. To answer the question what results in the harmonic ellipticity not only enables people to gain a deeper insight into the underlying physical mechanism of atomic HHG, but also help researchers understand more clearly about the generation of elliptically and circularly polarized molecular HH \cite{Zhou,Smirnova1,Etches2,Ramakrishna,Qin,Yuan}.

In a recent work, the polarization properties of the HH generated by atomic gases in elliptically polarized lasers was investigated and the ellipticity of the HH was attributed to the quantum-mechanical uncertainty of the electron motion \cite{Strelkov}. The authors have developed a theoretical model to describe the ellipticity and their analytical result agrees well with the exact numerical result for HH of high orders, but the analytical result is a little lower than the numerical one for low harmonic orders (see Fig. 2 in \cite{Strelkov}). In that work, the Coulomb effect is not taken into account and this effect is considered to be not responsible for the harmonic ellipticity at least for the short quantum path and not very low harmonic orders. In ref. \cite{Ramakrishna,Seideman,Zwan}, the authors have pointed out that the molecular potential is responsible for the observed ellipticity in molecular HHG, which inspires one that the deviation observed in ref. \cite{Strelkov} for low harmonic orders may result from the neglect of the Coulomb potential. In ref. \cite{Strelkov2}, the Coulomb effect was also introduced by using the continuum Coulomb state in the theoretical model for HHG in elliptically polarized field. However, in the previous works, how the potential affects the HHG and leads to the harmonic ellipticity is still not clear enough. Therefore, more detailed studies concerning the role of Coulomb effect on the generation of elliptically polarized HH are in demand.

In this paper, the effect of the Coulomb potential on the generation of elliptically polarized HH from atoms driven by intense elliptically polarized lasers is investigated analytically. By using the strong-field eikonal-Volkov approximation (SF-EVA), we in detail analyze the influence of the Coulomb potential on the dynamics of the continuum states and exposit the role of the Coulomb potential on the harmonic ellipticity. We further investigate the contribution of the Coulomb effect to the harmonic ellipticity for different harmonic orders and quantum paths, which explains the fact that the ellipticity is mainly obtained in the low order harmonics from short quantum path in our simulation. Moreover, the conclusions are further confirmed by comparing the results obtained with the soft-core potential and short-range potential.

\section{Theoretical model}
We employ 1300 nm, 2.2$\times$10$^{14}$ W/cm$^2$, elliptically polarized driving lasers with ellipticity of 0.1. The major axis of the polarization ellipse is parallel to the $x$ axis of the laboratory frame and the minor axis is parallel to the $y$ axis. To obtain the ellipticity of the HH, the time-dependent transition dipoles of the system are calculated by \cite{Levesque,Guhr}
\begin{equation}
d_{i}(t;k)=\langle\psi_0(\mathbf{r},t)|r_i|\psi_k(\mathbf{r},t)\rangle, i=x,y
\end{equation}
where $\psi_0$ is the ground state of the atom and $\psi_k$ refers to the continuum state. $\mathbf{k}$ is the recombination momentum, which is associated with the recombination time $t$.

In our simulation, the ground state $\psi_0$ is obtained by solving the time-dependent Schr\"{o}dinger equation (TDSE) with imaginary-time propagation. The Hamiltonian is (in atomic units)
\begin{equation}
H=-\frac{1}{2}\mathbf{\nabla}^2+V(\mathbf{r}),
\end{equation}
with
\begin{equation}
V(\mathbf{r})=-\frac{1}{\sqrt{r^2+a}},
\end{equation}
where the soft-core parameter $a$ is set to be 0.388 to fit the ground-state energy of Ar $E_g=-0.5794$ a.u. Through this paper, the target atom is Ar.

Regarding the continuum state $\psi_k$, if the continuum electron is treated as a free particle moving in the electric field without the effect of Coulomb potential, the continuum state is approximated by the plane wave $\psi_k(\mathbf{r},t)=e^{i\mathbf{k}(t)\cdot\mathbf{r}}$. In the expression, the space-independent terms for the continuum state $a_{ion}[t_b(t)]exp[-i\int_{t_b}^t{k_L^2(t')/2}dt']/(2\pi)^{(3/2)}$ are omitted, where $t_b$ is the ionization time, $\mathbf{k}_L(t')$ is the instantaneous momentum of the continuum electron in the laser field and $a_{ion}(t_b)$ refers to the ionization amplitude. These terms can be omitted because they will be canceled in calculating the amplitude ratio $R$ and the phase different $\delta$ between the two orthogonal components. The physical meaning of this treatment is that, according to the three step model, the HHG in the $x,y$ components share identical ionization and acceleration steps and the different properties between the two components of the HH radiation originate from the recombination step \cite{Levesque,Zhou,Itatani}.

For the plane waves, according to the Euler's formula $e^{i\mathbf{k}\cdot\mathbf{r}}=\cos(\mathbf{k}\cdot\mathbf{r})+i\sin(\mathbf{k}\cdot\mathbf{r})$, the real and imaginary parts of $\psi_k$ are symmetric and antisymmetric on the spatial coordinates respectively. As the ground state $\psi_0$ is also symmetric, when its space-independent term $e^{-iE_{g}t}$ is also omitted, the result of the integral $d_{i}(t;k)=\int\psi_{0}r_i\psi_{k}dr$ must be a pure imaginary number and the phase difference between the two components $d_y$ and $d_x$ is either 0 or $\pi$ \cite{Qin}. In the more realistic situation, when the space-independent terms are included, the result for the phase difference remains the same. So the calculated HH is linearly polarized and the harmonic ellipticity cannot be explained using the plane wave approximation due to the symmetry of the states \cite{Levesque}.

In this work, we employ the SF-EVA \cite{Smirnova2,Smirnova3,Bondar}, which enables one to describe the continuum dynamics taking into account the Coulomb effect. With this approach, the continuum state is expressed as
\begin{equation}
\psi_k(\mathbf{r},t)=e^{i\mathbf{k}(t)\cdot\mathbf{r}+i\sigma_k(\mathbf{r},t)}.
\end{equation}
$\sigma_k(\mathbf{r},t)$ is the correction to the plane wave considering the effect of the Coulomb potential and is given by
\begin{equation}
\sigma_k(\mathbf{r},t)=-\int_{t_b}^{t}{k_L(t')\{ k_L(t')-\sqrt{k_L^2(t')-2V[\mathbf{r}_L(t')]}\}}dt'£¬
\end{equation}
where
\begin{equation}
\mathbf{r}_L(t')=\mathbf{r}-\int_{t'}^{t}{\mathbf{k}_L(t'')}dt''
\end{equation}
describes the motion of the continuum electron in the laser field.

Compared with the continuum Coulomb state used in \cite{Strelkov2}, which only depends on the momentum $\mathbf{k}$ at time $t$, SF-EVA describes the Coulomb correction to the continuum state by tracing the motion of the electron in the process from $t_b$ to $t$. Therefore, with this approach, it is possible to in detail investigate influence of the Coulomb potential on the dynamics of the continuum electrons with different trajectories and to study the role of the Coulomb potential on the harmonic ellipticity. In ref. \cite{Smirnova3}, a detailed and systematic introduction to the SF-EVA is presented.

In this work, in the case of HHG driven by elliptically polarized lasers, we apply the same description for the motion of the continuum electron as in \cite{Strelkov}. In this description, the continuum electron starts from the origin with zero velocity and the return of the electron is judged by x=0. Therefore, the relation among $t$, $t_b$ and $\mathbf{k}$ can be obtained by finding the approximate solutions to the saddle-point equations \cite{Lewenstein,Antoine2,Becker,Chirila} in the x component. Then, the instantaneous momentum $\mathbf{k}_L(t)$ and the classical trajectory of the continuum electron can be calculated by substituting the solutions into the equations of the electron motion for both x and y components.

The amplitude ratio and the phase difference between the two orthogonal harmonic components are calculated by $R=|d_y|/|d_x|$ and $\delta=arg[d_y]-arg[d_x]$ respectively, and the ellipticity $\varepsilon$ is finally obtained by \cite{Son}
\begin{equation}
\varepsilon=\sqrt{\frac{1+R^2-\sqrt{1+2R^2cos2\delta+R^4}}{1+R^2+\sqrt{1+2R^2cos2\delta+R^4}}}.
\end{equation}
When the amplitude ratio $R=0$ or the phase difference $\delta=0$ or $\pi$, the ellipticity $\varepsilon=0$ and the HH is linearly polarized, otherwise it will be elliptically or circularly polarized. In this work, in order to study the role of Coulomb effect on the harmonic ellipticity, we only take into account the influence of the Coulomb potential in the model and the other effects such as the quantum mechanical uncertainty of the electron motion are not included.

\section{Result and discussion}
In Fig. 1, three typical trajectories of the continuum electrons with final recombination kinetic energy of $1U_p$, $2U_p$ and $3U_p$ are presented, where $U_p = F_0^2/4\omega^2 $ with $F_0$ and $\omega$ being the amplitude and angular frequency of laser. Panels (a) and (b) correspond to the short and long quantum paths respectively. The continuum electron starts from the origin with zero velocity and recombines to the core at $x=0$, which is the same as in ref. \cite{Strelkov}. As discussed above, these trajectories are obtained by solving the saddle-point equations in the x component and substituting the approximate solutions into the equations of the electron motion for both x and y components.

\begin{figure}[htb]
\centerline{
\includegraphics[width=12cm]{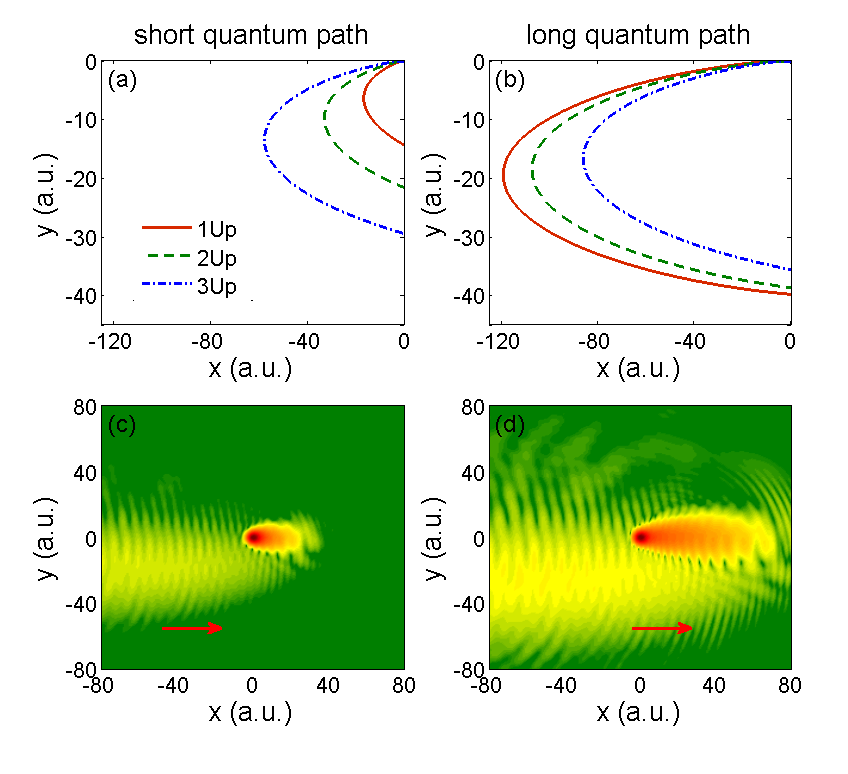}}
 \caption{(a) and (b) Typical trajectories of the continuum electrons with recombination kinetic energy of $1U_p$, $2U_p$ and $3U_p$ for short and long quantum paths respectively. (c) and (d) Snap shots of the evolution of the probability density distribution of the wavepacket at the instant of recombination.}
\end{figure}

According to the cutoff law of HHG \cite{Lewenstein}, the recombination kinetic energy of $3U_p$ corresponds to HH radiation at the cutoff in the harmonic spectrum, and the recombination kinetic energy of $2U_p$, $1U_p$ correspond to HH of lower orders in the plateau region (the 89th and 53rd orders respectively). As shown in Fig. 1(a) for short quantum path, when the recombination kinetic energy grows higher, the trajectory grows longer and the motion of the continuum electron is farther from the core. The maximum distance increases from no more than 20 a.u. to nearly 60 a.u. As shown in Fig. 1(b) for long quantum path, the trajectories are even longer for HH from the cutoff to lower orders. At the same time, with the trajectory extending longer, the transverse shift grows up to 40 a.u. Figs. 1(c) and (d) are the snap shots of the evolution of the probability density distribution of the wavepacket at the instant of recombination. The evolution is obtained by solving the TDSE numerically. In the quantum view of the dynamic of the continuum electron, one can see that although the center of the continuum wavepacket shifts about 30 a.u. transversely, it is still wide enough to interfere with the bound state to generate the HH. For the numerical evolution, the laser parameters are the same as those given in Section 2 and the initial state is also obtained by solving TDSE with imaginary-time propagation with the Hamiltonian given by Eq. (2).

\begin{figure}[htb]
\centerline{
\includegraphics[width=12cm]{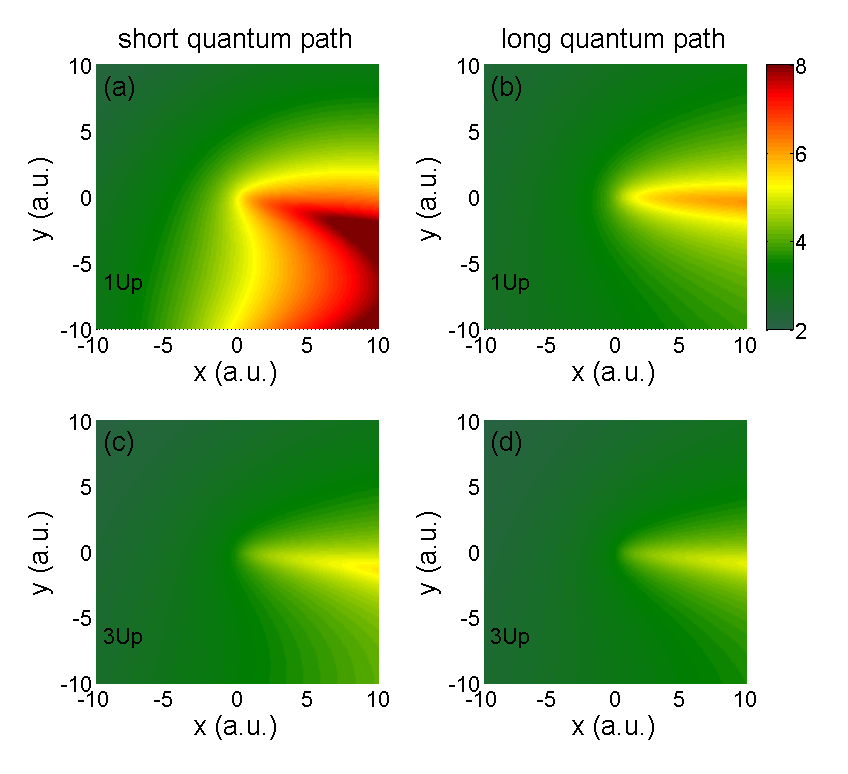}}
\caption{Calculations of the Coulomb modification term $\sigma_k(\mathbf{r},t)$ at the instant of recombination for (a) short quantum path and low harmonic order, (b) long quantum path and low harmonic order, (c) short quantum path and high harmonic order and (d) long quantum path and high harmonic order. }
\end{figure}

It has been shown in Fig. 1 that the behaviors of the continuum electrons are different among different harmonic orders and quantum paths. These different features of the motions of the continuum electrons will result in the different harmonic ellipticities in the HHG for different harmonic orders and quantum paths. According to the previous discussion, if the continuum state is approximated by the plane wave, the obtained HH will be linearly polarized. However, when we take into account the Coulomb effect, during the acceleration step of HHG, the continuum state will be modified by the Coulomb potential before recombination. This modification is expressed by the term $\sigma_k(\mathbf{r},t)$ in Eq. (5), which is mainly determined by the integral for the Coulomb potential along the trajectories, \emph{i.e.} $V[\mathbf{r}_L(t')]$. After considering this modification in the phase, the continuum state will be no longer a plane wave. As a result, the symmetry of the wave function $\psi_k$ is broken and the obtained $d_x$, $d_y$ are not pure imaginary numbers any more. Then, the value of the phase difference $\delta$ varies in the interval of [0,2$\pi$) rather than fixes to 0 or $\pi$ \cite{Qin}. Therefore, the Coulomb effect on the continuum states finally leads to the elliptically polarized HHG.

In Fig. 2, we calculate the modification term $\sigma_k(\mathbf{r},t)$ at the instant of recombination with kinetic energy of $1U_p$ and $3U_p$ and for short and long quantum paths respectively. All the four panels are plotted in the same color scale. Comparing the second column to the first column in Fig. 2, the Coulomb effect is weaker for long quantum path. This is because for the long quantum path, the electron leaves directly far away from the core, drifts in the faraway region where the Coulomb potential $V(\mathbf{r})$ is weak and enters the vicinity of the core shortly before recombination. While for short quantum path, the electron only drifts in the vicinity of the core, where the effect of the Coulomb potential is much stronger.

In addition, comparing the first row to the second row, for both the short and long quantum paths, the Coulomb effect for lower harmonic orders is stronger than that for higher orders. This is because, in the case of short quantum path, the trajectory of the continuum electron for the lower harmonic order is closer to the core and the influence of the Coulomb potential $V[\mathbf{r}_L(t)]$ is much bigger. Thus it is obvious that the Coulomb effect is strongest for the low order harmonics from short quantum path as shown in panel (a). For the long quantum path, although the trajectory for $1U_p$ is longer and farther from the core than that for $3U_p$, the Coulomb effect is already very weak for both faraway electrons. The determining factor is that, the accumulated Coulomb phase is greater for the slow electron especially in the processes when the electron leaves and comes back to the vicinity of the core. Mathematically speaking, the integral between $t_b$ and $t$ in Eq. (5) is bigger for HH of low harmonic order.

Finally the calculated rotation angle of the HH field and the ellipticity as functions of the recombination time $t$ are presented in Fig 3. The recombination at about $t=0.7$ optical cycle is associated with the HH at the cutoff region in the harmonic spectrum, and recombination at earlier or later time corresponds to HH of lower harmonic orders in the plateau region. The rotation angle is defined by the direction of the electron momentum at the instant of return \cite{Strelkov}. Nearly identical curve is observed as that in ref. \cite{Strelkov}, which indicates that our description to the motion of the continuum electron is the same. Regarding the ellipticity, as shown by the solid blue curve, unnegligible harmonic ellipticity is observed. The ellipticity is up to more than 0.2 for HH of low harmonic orders contributed by short quantum path. The result shows that the Coulomb effect also makes a contribution to the harmonic ellipticity at least for the low harmonic orders and short quantum path.

\begin{figure}[htb]
\centerline{
\includegraphics[width=10cm]{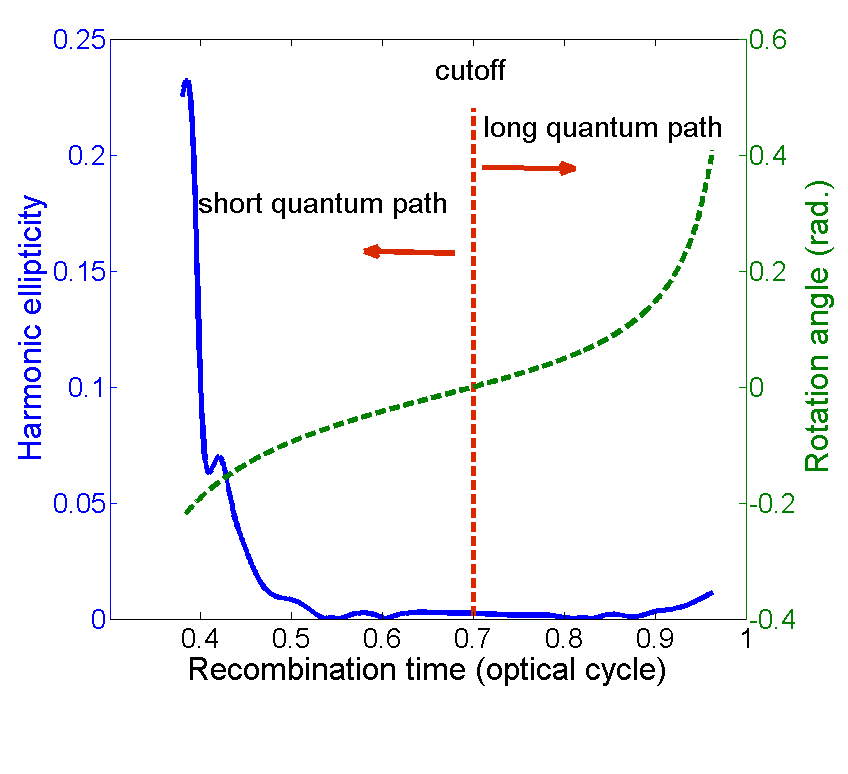}}
\caption{The harmonic ellipticity and the rotation angle as functions of the recombination time $t$.}
\end{figure}

When we transform the ellipticity curve as a function of the recombination time $t$ into curves as a function of the harmonic orders, it will be more obvious for the result to be analyzed. This is shown in Fig 4(a) and the phase differences between the two orthogonal components $\delta$ are also plotted in Fig 4(b). As the relationship between $t$ and $\mathbf{k}$ has been obtained by solving the saddle-point equations in the x component, the recombination momentum is then transformed into the harmonic order $q$ according to the relation $q\omega=k^2/2-E_g$.

It is found in Fig. 4(a) that the harmonic ellipticity of short quantum path is much higher than that of long quantum path, and that the ellipticity for both quantum paths drops as the harmonic order increases. According to the previous discussions and analyses, the obtained ellipticity originates from the break of symmetry of the continuum state and the break of symmetry results from the Coulomb modification. In Fig. 2, it is shown that the Coulomb modification for the short quantum path is much stronger than that for the long quantum path, and the Coulomb modification for low harmonic orders is stronger than that for high harmonic orders. Therefore, all the observations above are consistent with the previous discussions and can be well explained by the analyses of Fig. 2. The vanishment of the ellipticity for the high harmonic orders implies that the effect of the Coulomb potential on high energy electrons is much weaker and the Coulomb continuum state is reduced to approximate plane waves, which also agrees with the prediction in \cite{Strelkov}.

Comparing Fig. 4(b) with Fig. 4(a), the relationship between the ellipticity and the phase difference is also consistent with the previous discussion. For high harmonic orders, the ellipticity falls to zero when the phase difference equals the value of $\pi$ due to the symmetry of the ground state and the continuum state $\psi_k\approx e^{i\mathbf{k}\cdot\mathbf{r}}$. For low harmonic orders, where the Coulomb effect is strong enough to modify the continuum state, $\psi_k$ is no longer a plane wave and the symmetry is broken. As a result, the phase different deviate from $\pi$ and the nonzero ellipticity appears.

To find in the frequency region where the Coulomb effect is unnegligible, we also compare our results with the analytical results in ref. \cite{Strelkov} (also presented by the green curves in Fig. 4(a)). For the harmonics in the range from the 21st to the 61st orders, the ellipticity of short quantum path obtained in our work is considerable and is in the same order of magnitude compared with that in \cite{Strelkov}. Note that in ref. \cite{Strelkov}, it is also in the same range from the 21st to the 61st harmonic order where their analytical result is lower than the numerical result. For harmonics higher than the 61st order, the ellipticity obtained in our work rapidly decreases to zero, and meanwhile the analytical results agrees well with the numerical results in ref. \cite{Strelkov}. This comparison shows that, the Coulomb effect also contributes to the ellipticity for the low order harmonics, and neglecting the Coulomb effect would lead to the deviation of the result. For the long quantum path, our result is nearly negligible for all the harmonic orders compared with that in \cite{Strelkov}. This is because the Coulomb effect is already very low for the long quantum path, where the continuum electron drifts far away from the core.

\begin{figure}[htb]
\centerline{
\includegraphics[width=15cm]{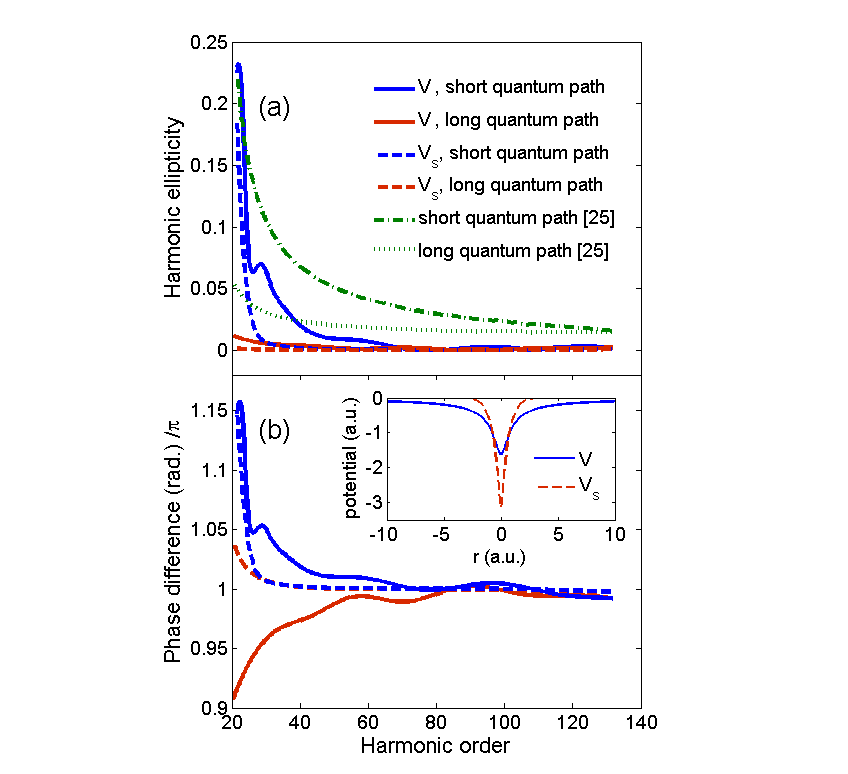}}
\caption{(a) The harmonic ellipticity and (b) the phase difference between the two orthogonal components as functions of the harmonic order. Inset: Comparison between the soft-core potential $V(\mathbf{r})$ and the short-range potential $V_s(\mathbf{r})$. The analytical results in \cite{Strelkov} are also presented in panel (a) with the green curves.}
\end{figure}

To further investigate the role of Coulomb potential on the harmonic ellipticity, we also calculate the ellipticity with short-range potential $V_s(\mathbf{r})=-1/\sqrt{r^2+a}\exp(-r^2/s^2)$ \cite{Smirnova2}, where s=1.6 and a=0.102 to keep the ground-state energy unchanged. As shown in the inset of Fig. 4, the short-range potential approaches zero much closer to the core at about r=3 a.u. The obtained ellipticity with short-range potential is shown in Fig. 4(a) with dashed curves. Compared with the solid curves, the dashed curves are even sharper and drop to zero more quickly around the 31th harmonic order. This is because the potential vanishes so rapidly that the trajectory of the associated continuum electron is already too far away for the short range of $V_s(\mathbf{r})$. In this relatively faraway area, the Coulomb effect is too weak to lead to the elliptically polarized HHG. However, for the HH of the lowest orders of short quantum path, the motions of the continuum electrons are confined in a very small range. Take the 21st harmonic order for example, the maximum distance of the electron trajectory from the core is only about 2.5 a.u. In this area, the difference of the Coulomb effects on these trajectories between the short- and long-range potentials is small. Therefore, the harmonic ellipticity of these orders is still high and become close to that obtained with soft-core potential. Similarly, as shown by the dashed curves in Fig. 4(b), the phase differences also converge to the value of $\pi$ much more rapidly due to the short range of the Coulomb potential.

\section{Conclusion}
In this paper, the role of the Coulomb potential on the generation of elliptically polarized HH from atoms driven by elliptically polarized lasers has been investigated analytically. It is found that the Coulomb effect makes a contribution to the harmonic ellipticity for low harmonic orders and short quantum path. We in detail analyze the influence of the Coulomb potential on the dynamics of the continuum state and find that the obtained ellipticity in our simulation originates from the break of symmetry of the continuum state. Moreover, the conclusions are further confirmed by comparing the results obtained with the soft-core potential and the short-range potential. We would like to further mention that this kind of Coulomb potential induced ellipticity is general for different species of core. Therefore, this mechanism can also, at least in part, result in the harmonic ellipticity in molecular HHG. In addition, with the analytical model in SF-EVA, we have not only proposed the mechanism of the Coulomb effect on the harmonic ellipticity but also depicted the electronic dynamics of the Coulomb continuum states, which would be helpful for many other different topics in strong-field phenomena, such as finer molecular orbital tomography and ultrafast dynamic tracing.

\section*{Acknowledgment}
This work was supported by the National Natural Science Foundation
of China under Grants No. 60925021, 10904045, the National
Basic Research Program of China under Grant No. 2011CB808103, and the Doctoral fund of Ministry of Education of China under Grant No. 20100142110047. This
work was partially supported by the State Key Laboratory of
Precision Spectroscopy of East China Normal University.
\end{document}